%% file: Arxiv.tex
\documentclass[conference]{IEEEtran}
\IEEEoverridecommandlockouts

\usepackage{cite}
\usepackage{amsmath,amssymb,amsfonts}
\usepackage{algorithmic}
\usepackage{graphicx}
\usepackage{textcomp}
\usepackage{xcolor}
\usepackage{url}
\usepackage{multirow}
\usepackage{tabularx}
\usepackage{subcaption}
\usepackage{xcolor}
\usepackage{tikz}
\usetikzlibrary{arrows.meta, positioning, shapes.geometric}

\def\BibTeX{{\rm B\kern-.05em{\sc i\kern-.025em b}\kern-.08em
    T\kern-.1667em\lower.7ex\hbox{E}\kern-.125emX}}

\begin{document}

\title{Measurement-Driven Early Warning for Resilient 5G NSA Railway Networks \vspace{-0.1in}
\thanks{This work was supported in part by the National Science and Technology Council (NSTC) of Taiwan under Grant 113-2926-I-001-502-G. Prof. Walid Saad was supported by the U.S. National Science Foundation (NSF) under Grant CNS-2114267.}
}

\author{
\IEEEauthorblockN{Po-Heng Chou$^{1,3}$, Da-Chih Lin$^{2}$, Hung-Yu Wei$^{2}$, Walid Saad$^{3}$, and Yu Tsao$^{1}$}
\IEEEauthorblockA{
$^{1}$Research Center for Information Technology Innovation (CITI), Academia Sinica (AS), Taipei 11529, Taiwan\\
$^{2}$Department of Electrical Engineering, National Taiwan University (NTU), Taipei 10617, Taiwan\\
$^{3}$Bradley Department of Electrical and Computer Engineering (ECE), Virginia Tech (VT), Alexandria, VA 22305, USA\\
E-mails: d00942015@ntu.edu.tw, f13921042@ntu.edu.tw, hywei@ntu.edu.tw, walids@vt.edu, yu.tsao@citi.sinica.edu.tw
\vspace{-0.2in}}}

\maketitle

\begin{abstract}
This paper presents a measurement-driven benchmark for seconds-ahead early warning of reliability breakdown events in commercial fifth-generation (5G) non-standalone (NSA) railway networks. Using 10~Hz metro-train traces containing serving- and neighboring-cell measurements, we formulate the prediction of Master Cell Group Failure (MCGF)- and Non-Access Stratum Recovery (NASR)-related events as a supervised time-series classification problem. Six representative models, including convolutional neural network (CNN),
long short-term memory (LSTM), XGBoost, Anomaly Transformer, PatchTST, and TimesNet, are evaluated under multiple observation windows and prediction horizons. TimesNet achieves the highest F1 score of 0.8498 with a 3-s observation window and a 3-s prediction horizon, while CNN achieves a competitive F1 score of 0.8208 with a 3-s observation window and a shorter 2-s prediction horizon. Event-oriented analysis further shows that less restrictive trigger rules provide high event coverage, whereas consecutive-decision rules enforce more persistent triggers at the cost of missed events. The results demonstrate the predictive value of lightweight device-side radio measurements for proactive resilience in the evaluated 5G NSA metro environment.
\end{abstract}


\section{Introduction}

The fifth-generation (5G) wireless network has become a key enabler for emerging mission-critical applications that demand highly reliable communications~\cite{Haque2025_URLLC}. Among the vertical sectors benefiting from 5G, railway communications represent one of the most reliability-sensitive domains, as wireless links support train control, operational coordination, and information exchange between onboard and ground systems~\cite{Jiao2022_TrainControl,Nikolopoulou2022_5GRAIL}. Any delay, packet loss, or temporary interruption of such communication may affect service continuity, traffic efficiency, and, in safety-critical cases, passenger safety~\cite{Jiao2022_TrainControl}.
Therefore, improving the resilience of 5G-based railway communications requires not only rapid recovery after a disruption but also the capability to anticipate reliability degradation before an observable service breakdown occurs~\cite{Nikolopoulou2022_5GRAIL, Bi2026_Resilience}.

Commercial 5G non-standalone (NSA) networks introduce additional mobility-management complexity through dual connectivity between Long-Term Evolution (LTE) evolved Node Bs (eNBs) and 5G New Radio (NR) next-generation Node Bs (gNBs). Under high-speed train mobility, frequent changes in serving and secondary cells may trigger repeated reconfiguration procedures, signaling bursts, transient link degradation, or temporary disconnection. These effects are particularly relevant in dense railway environments, where overlapping cells, rapid channel variation, and limited reaction time make purely reactive reliability mechanisms insufficient for some service-critical situations.

Despite the rapid deployment of commercial 5G NSA networks, empirical studies have shown that their real-world reliability and latency performance can differ substantially from idealized expectations. Measurements across urban environments have reported non-negligible packet loss and latency fluctuations when compared with mature LTE networks~\cite{Xu2020_SIGCOMM5G}. Such discrepancies become more pronounced in high-mobility scenarios, where frequent cell handovers may lead to transient disconnections and unstable throughput. Our previous field experiments along the Taipei Metro confirmed this phenomenon: packet-loss and excessive-latency events were frequently observed near station areas with intensive handover activity~\cite{Lin2022_VTC5GMetro}. By analyzing lower-layer signaling messages captured using MobileInsight~\cite{Li2016_MobiComMobileInsight}, we further observed that handover-related events constitute a major source of reliability degradation in the measured 5G NSA railway environment.

To mitigate these reliability issues, several studies have investigated device-side and network-side mechanisms for improving mobile connectivity. Representative approaches include multi-connectivity, redundant transmission, mobility enhancement, and recovery-oriented mechanisms that maintain alternative links, reduce unnecessary handovers, or restore service after degradation has emerged~\cite{Liu2024_M2HO,An2023_Octopus}. These approaches are important for maintaining service continuity and supporting post-event recovery. However, they are primarily reactive because mitigation is generally activated only after degradation symptoms or service disruption have become observable.

From a network-resilience perspective, an additional early-warning capability is therefore desirable. Predicting an impending reliability breakdown can provide a short preparation interval for redundancy activation, handover assistance, traffic adaptation, or other preventive actions before the disruption becomes severe. Such predictive warning does not replace existing recovery mechanisms. Instead, it complements them by extending network resilience from reactive recovery toward proactive anticipation and mitigation readiness.

Learning-based approaches have recently been applied to radio link failure (RLF) and communication-failure prediction. Existing studies have shown that impending failures can be inferred from key performance indicators, control-plane signaling, and temporal network observations~\cite{Islam2023_TNSM_DNNRLF}. More recent works have investigated transformer-based models~\cite{Farooq2025_ICC_TransformerRLF} and graph-enhanced architectures~\cite{Hasan2025_TMLCN_GNNTransformerRLF} to capture temporal and spatial dependencies in cellular networks. However, existing studies often rely on network-side KPIs, contextual data, or non-railway evaluation settings, which may not fully reflect device-side measurements under operational railway mobility.

In contrast, this paper presents a measurement-driven study of seconds-ahead early warning using field-collected measurements from a commercial 5G NSA metro environment. Rather than proposing a new prediction architecture, we investigate whether lightweight radio measurements available from commercial devices contain sufficient predictive information to anticipate impending reliability breakdown. The objective is to quantify the feasibility and operating trade-offs of learning-based warning under authentic high-mobility, interference, and signaling conditions.

The prediction target is defined using practically observable stack-level events, primarily Master Cell Group Failure (MCGF) and Non-Access Stratum Recovery (NASR). These events are associated with substantial packet loss and service interruption in the measured 5G NSA network. They should not, however, be interpreted as being identical to the formal Third Generation Partnership Project (3GPP) RLF procedure based on out-of-sync indications and timer expiry. Accordingly, this paper refers to them as \emph{RLF-related reliability breakdown events}.

Using 10~Hz field-collected measurements containing reference signal received power (RSRP), reference signal received quality (RSRQ), and serving- and neighboring-cell information, we formulate early warning as predicting whether an RLF-related event will occur within horizon $T_p$ from an observation window of duration $T_s$. We benchmark convolutional neural network (CNN)~\cite{LeCun2015_NatureDL}, long short-term memory (LSTM)~\cite{Hochreiter1997_LSTM}, XGBoost~\cite{Chen2016_XGBoost}, Anomaly Transformer~\cite{Xu2022_AnomalyTransformer}, PatchTST~\cite{Nie2023_PatchTST}, and TimesNet~\cite{Wu2023_TimesNet}.

From the perspective of resilient network operation, the warning output can support proactive actions such as redundancy preparation, handover assistance, or traffic adaptation before severe degradation occurs. This study focuses on warning feasibility and operating trade-offs; closed-loop control design is left for future work.

The main contributions of this paper are summarized as follows:

\begin{itemize}

\item \textbf{Measurement-driven characterization of reliability breakdown:}
Using field measurements from a commercial 5G NSA metro environment, we identify MCGF and NASR as practically important RLF-related events and characterize their relationship with packet-loss behavior under high mobility.

\item \textbf{Benchmarking of seconds-ahead early warning:}
We formulate reliability-breakdown warning as a supervised time-series classification problem and compare six representative learning models under multiple observation windows and prediction horizons.

\item \textbf{Resilience-oriented operational analysis:}
We examine the trade-offs among warning horizon, prediction performance, computational cost, and event-oriented trigger behavior, providing preliminary guidance for integrating predictive warning with proactive mobility and reliability-control mechanisms.

\end{itemize}

\begin{figure*}[t]
\centering
\includegraphics[width=2.0\columnwidth]{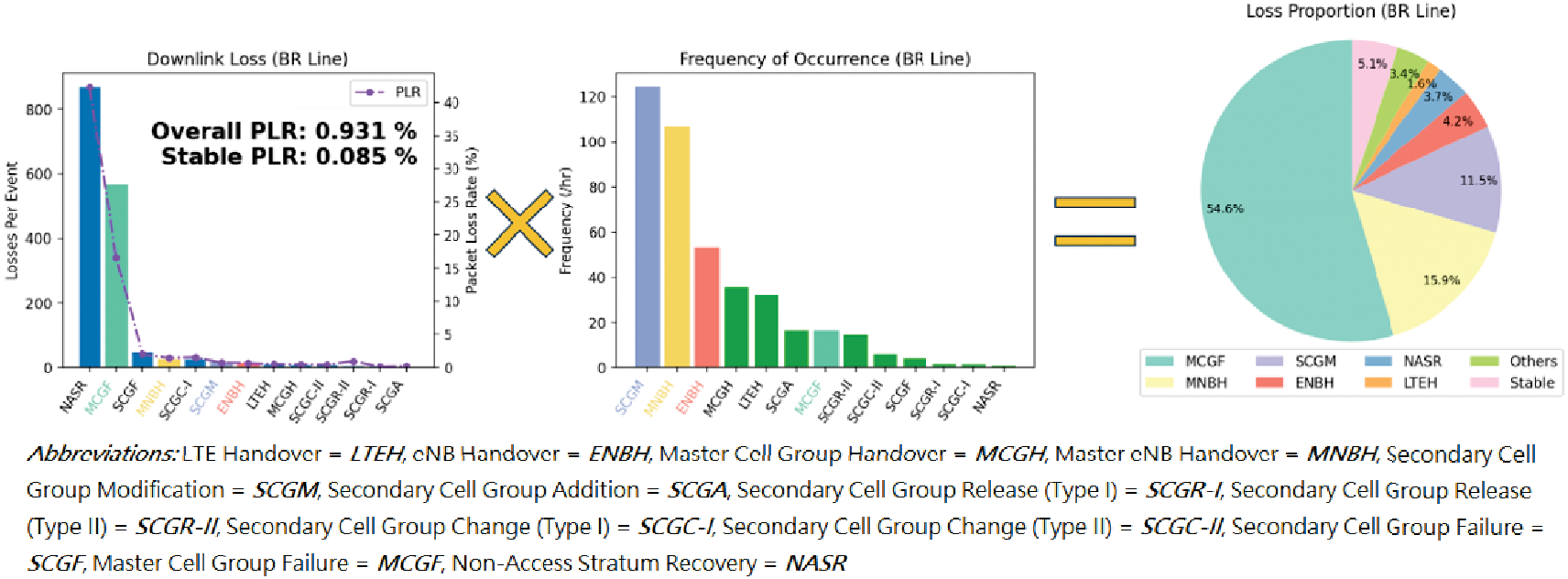}
\caption{Downlink packet loss distribution by event types and occurrence frequency.}
\label{fig1}
\vspace{-0.2in}
\end{figure*}

\section{Related Works}

Related studies fall into three categories: measurement-based reliability analysis, reliability enhancement, and learning-based failure prediction.

\subsection{Measurement-Based Reliability Studies}

Measurement campaigns have shown that the reliability and latency performance of commercial 5G networks can differ from idealized expectations. Compared with mature LTE systems, operational 5G deployments may exhibit non-negligible packet loss, latency variation, and unstable performance under mobility~\cite{Xu2020_SIGCOMM5G}. These effects are particularly relevant in railway environments, where rapid movement and frequent cell transitions increase the likelihood of transient communication degradation.

Our previous measurements on the Taipei mass rapid transit (MRT) system examined 5G NSA reliability and latency under real train operation~\cite{Lin2022_VTC5GMetro}. The results showed that packet-loss and excessive-latency events frequently occurred near metro stations with intensive handover activity. Statistical analysis further indicated that up to $96\%$ of downlink packet-loss events were observed during handover-related intervals. These findings motivate the present work, which moves beyond post-event measurement analysis and investigates whether impending reliability breakdown can be anticipated from radio measurements collected before the event.

\subsection{Reliability Enhancement and Recovery Mechanisms}

Several studies have investigated mechanisms for improving mobile communication reliability or mitigating performance degradation. Early measurement-driven work studied the predictability of mobile bandwidth~\cite{Yao2008_Bandwidth}, while RAVEN used redundant connectivity to improve interactive latency in connected-vehicle scenarios~\cite{Lee2018_RAVEN}. More recent studies have investigated handover mitigation, edge-assisted performance enhancement, and artificial intelligence (AI)-based proactive mobility management~\cite{Liu2024_M2HO,An2023_Octopus,Li2025_AIMobility}. In particular, AI-based handover-event prediction has been considered for initiating cell-level mobility actions before signal degradation becomes severe in high-speed railway environments~\cite{Li2025_AIMobility}.

These approaches provide important mechanisms for maintaining service continuity, reducing handover impact, or recovering from degraded links. However, their operation is mainly activated after network degradation becomes observable or when a mobility event is already underway. In contrast, the present work focuses on a complementary early-warning capability that estimates whether an RLF-related reliability breakdown will occur within a future prediction horizon. Such warning can provide additional preparation time for redundancy activation, handover assistance, or other proactive resilience mechanisms.

\subsection{Learning-Based Failure Prediction}

Machine learning has increasingly been applied to fault prediction and anomaly detection in communication networks. A recent survey reviews learning-based fault-prediction methods ranging from conventional neural models to graph- and transformer-based architectures~\cite{Murphy2024_TNSM_FaultSurvey}. In 5G radio access networks, Islam et al.~\cite{Islam2023_TNSM_DNNRLF} demonstrated communication-failure prediction using key performance indicators and control-plane information. Transformer-based link-failure detection was subsequently investigated in~\cite{Farooq2025_ICC_TransformerRLF}, while a generalized graph neural network (GNN)--Transformer framework was proposed in~\cite{Hasan2025_TMLCN_GNNTransformerRLF} to capture temporal and spatial dependencies. Learning-based anomaly detection has also been studied in broader 5G Internet of Things (IoT) environments using hybrid CNN--LSTM models~\cite{Pirbhulal2025_ICC_AD5GIoT}.

Unlike studies centered on a particular prediction architecture, this work does not propose a new learning model. Instead, it provides a measurement-driven benchmark of six representative time-series models using traces collected from a commercial 5G NSA metro network. The emphasis is on evaluating the feasibility and operating trade-offs of seconds-ahead early warning under real mobility, signaling, and radio-measurement conditions. The resulting analysis connects learning-based prediction with proactive resilience by examining observation windows, prediction horizons, computational cost, and event-oriented trigger behavior.

\section{Measurements and Observations}

\subsection{Experimental Environment and Tools}
This work builds upon the experimental framework established in our previous measurement campaigns~\cite{Lin2022_VTC5GMetro}, employing the same devices, software tools, and test route to ensure data consistency and comparability. All experiments were conducted along the Taipei MRT Brown Line, from Xinhai Station to Taipei Zoo Station, a route that spans urban, semi-open, and underground segments.

In each experiment, the user equipment (UE) was implemented using commercial smartphones connected via universal serial bus (USB) to a laptop, which served as the data client. A remote server located at National Taiwan University acted as the communication endpoint. The downlink and uplink data streams were transmitted using the user datagram protocol (UDP), with a fixed payload size of 250~bytes and a transmission rate of 200~kbps.

A custom Android monitoring application was developed to collect key measurement parameters in real time. The tool sampled radio-layer indicators at 10~Hz, corresponding to ten measurement points per second, including the RSRP and RSRQ of both serving and neighboring LTE and 5G cells. These physical-layer metrics were synchronized with global positioning system (GPS) coordinates, enabling spatio-temporal correlation between network performance and geographical location.

For control-plane observation, we employed MobileInsight~\cite{Li2016_MobiComMobileInsight}, an open-source tool that extracts LTE and 5G NR signaling events from modem logs. This allowed us to record detailed radio resource control (RRC) procedures, such as connection setup, reconfiguration, and handover events. The integration of transport-layer packet traces with low-layer signaling messages enabled a comprehensive view of how physical and control-layer behaviors jointly influence communication reliability in high-mobility 5G NSA networks.
Although limited to one metro line and operator, the route spans urban, semi-open, and underground segments.

\subsection{Observations}
\label{sec:measure_observation}

Fig.~\ref{fig1} summarizes the downlink packet loss rate (PLR) and event occurrence statistics observed in our experiments. The analysis indicates that two specific events, NASR and MCGF, exhibit the highest packet loss per event. When the overall frequency and impact are combined, the five most influential events are MCGF, \textit{Master eNB Handover} (MNBH), \textit{Secondary Cell Group Modification} (SCGM), \textit{eNB Handover} (ENBH), and NASR, in descending order of contribution.

A critical pattern emerges when correlating these events with radio link reliability: both MCGF and NASR correspond to reliability breakdown events that are strongly associated with radio link failure behavior in the measured 5G NSA network. Together, they account for 58.3\% of all downlink packet losses, with MCGF alone responsible for over half of the total (54.6\%). This finding confirms that RLF-related events are the dominant source of reliability degradation in the measured 5G NSA railway environment.

These observations indicate that many reliability breakdown events are preceded by measurable radio-layer fluctuations, suggesting that lightweight device-side indicators can support early warning before a full breakdown. Developing learning-based early warning is therefore essential for improving the reliability and resilience of 5G railway communications.

\begin{figure}[t]
\centering
\includegraphics[width=0.85\columnwidth]{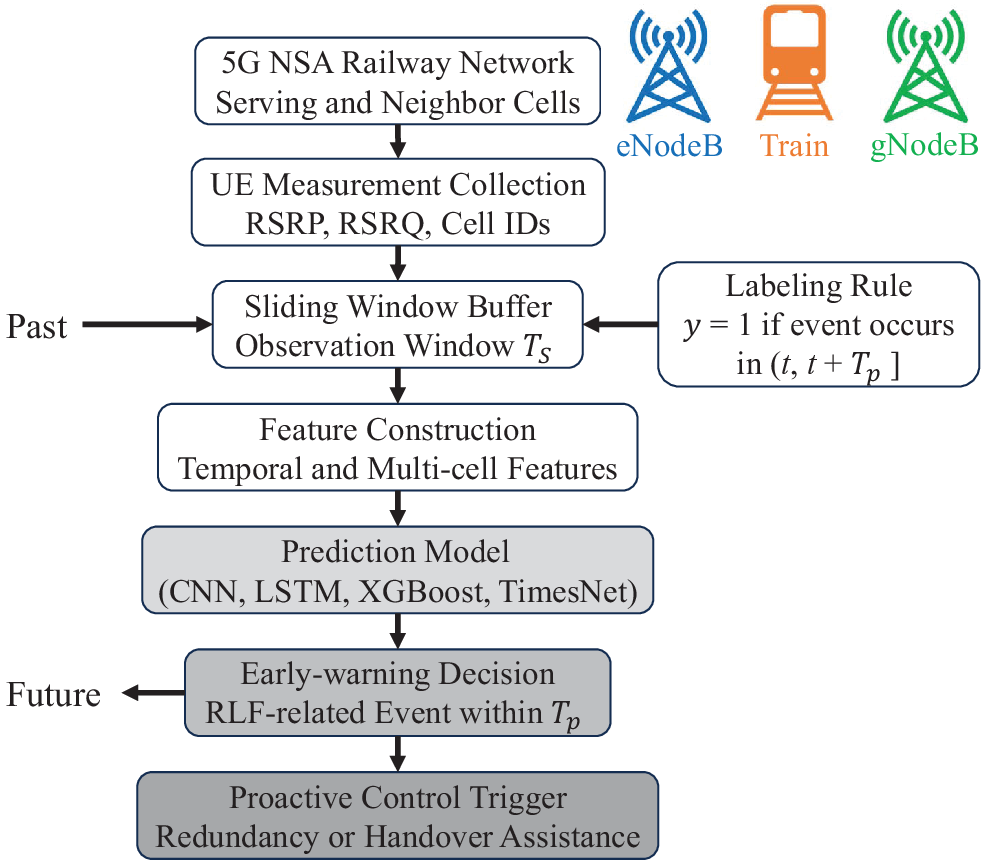}
\caption{System overview of the proposed measurement-driven early-warning framework.}
\label{fig0}
\vspace{-0.2in}
\end{figure}

Fig.~\ref{fig0} illustrates the system-level workflow of the measurement-driven early-warning framework. The UE continuously collects radio measurements from serving and neighboring cells, including RSRP, RSRQ, and cell identifiers. At each decision time $t$, these measurements are organized into a sliding observation window of length $T_s$ and converted into temporal multi-cell features. A prediction model then estimates whether an RLF-related reliability breakdown event will occur within the future horizon $(t,t+T_p]$. The resulting warning output can serve as a trigger for proactive mobility control, such as redundancy preparation, handover assistance, or reliability-aware alarm generation.

In this paper, the positive class is defined using stack-level events observed from operational logs, primarily MCGF and NASR. MCGF accounts for the dominant fraction of packet loss among the observed breakdown events, whereas NASR represents an additional recovery-related manifestation of severe service disruption in the commercial NSA protocol stack. We therefore consider both event types to avoid restricting the warning target to a single stack-level failure mechanism. These events do not strictly correspond to the formal 3GPP RLF definition based on out-of-sync indications and timer expiry. Instead, they represent practically observable reliability breakdown events associated with packet loss or service interruption in commercial 5G NSA operation. Therefore, the prediction target in this work is more precisely interpreted as RLF-related events.
To proactively address RLF events, we formulate the task as a supervised time-series classification problem. Let $\boldsymbol{X}_t$ denote the sequence of radio measurements collected within the observation window ending at time $t$, and let $\hat{y}_t$ denote the predicted probability of an RLF-related event occurring within $(t,t+T_p]$. A decision threshold of 0.5 is used to convert $\hat{y}_t$ into a binary warning decision. At each decision time $t$, the predictor takes as input a sliding observation window of duration $T_s$, consisting of historical measurements sampled at an interval of $I_s$.
For supervised training, the binary label $y\in\{0,1\}$ is set to one if at least one RLF-related event occurs within $(t,t+T_p]$, and to zero otherwise.

\subsubsection*{Feature Selection}
To capture the comprehensive characteristics of radio link quality, the RLF predictor employs a diverse set of features. The primary inputs include direct signal strength indicators such as RSRP and RSRQ. In addition, protocol-level information derived from RRC measurements and event reports provides critical context on network state transitions. To address spatial variability in railway environments, the feature set is further expanded to include the RSRP, RSRQ, and cell identifiers of the first three neighboring cells reported in each measurement interval, enabling the anticipation of handovers and the detection of coverage deficiencies.

\subsubsection*{Handling Class Imbalance}
With $I_s=0.1$~s (10~Hz), the positive-to-negative ratio is approximately 1:500. To mitigate this imbalance, class-weighted loss functions were used for the neural models, while XGBoost was configured with
$\textbf{scale\_pos\_weight}=\frac{\textbf{n\_neg}}{\max(1,\textbf{n\_pos})}$,
where $\textbf{n\_neg}$ and $\textbf{n\_pos}$ denote the numbers of negative and positive training samples, respectively.

\section{Performance Results}

\begin{figure*}[t!]
\centering
\begin{subfigure}{0.31\textwidth}
\centering
\includegraphics[width=\textwidth]{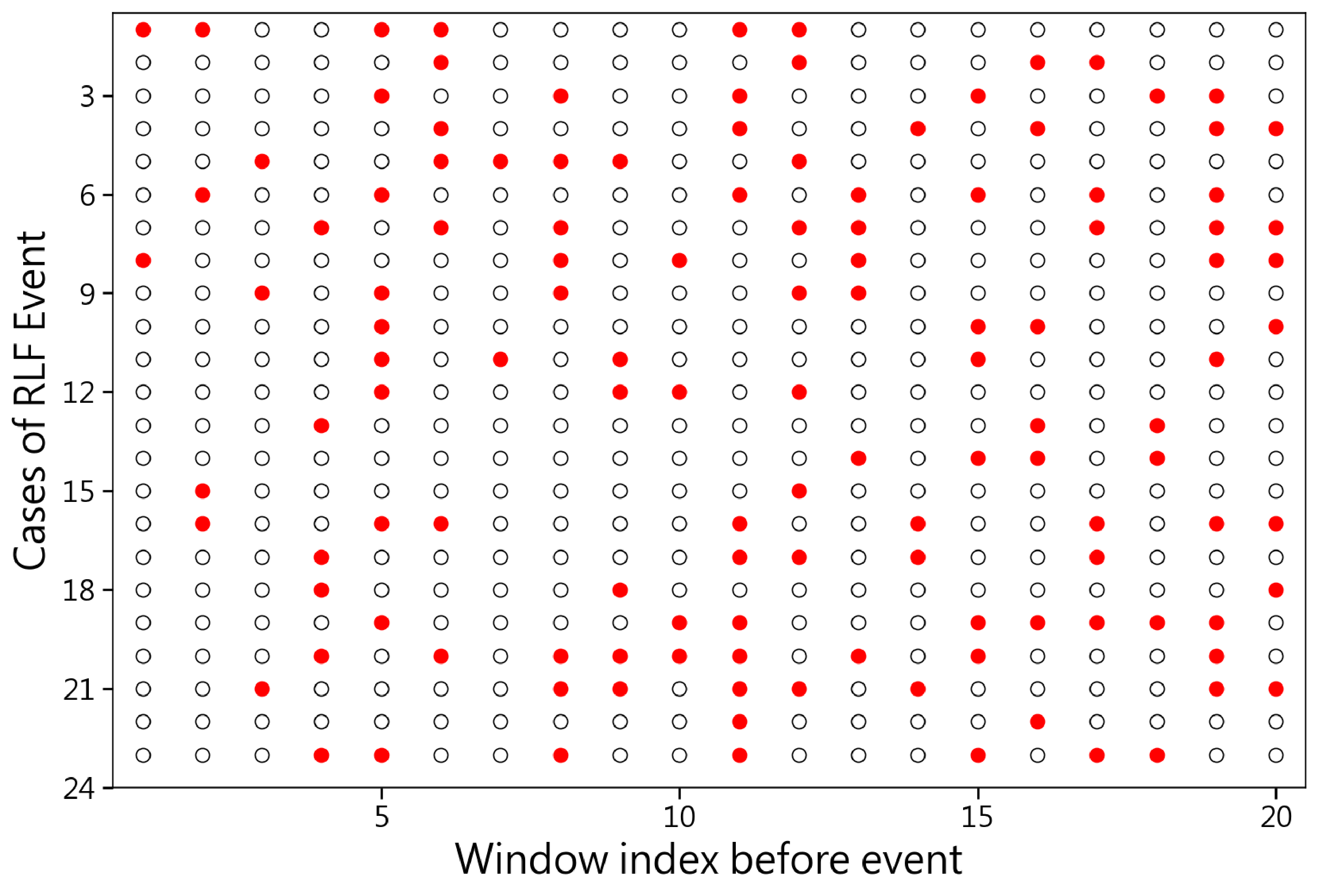}
\caption{CNN~\cite{LeCun2015_NatureDL} with $T_p=2$~s}
\label{fig4a}
\end{subfigure}
\hfill
\begin{subfigure}{0.31\textwidth}
\centering
\includegraphics[width=\textwidth]{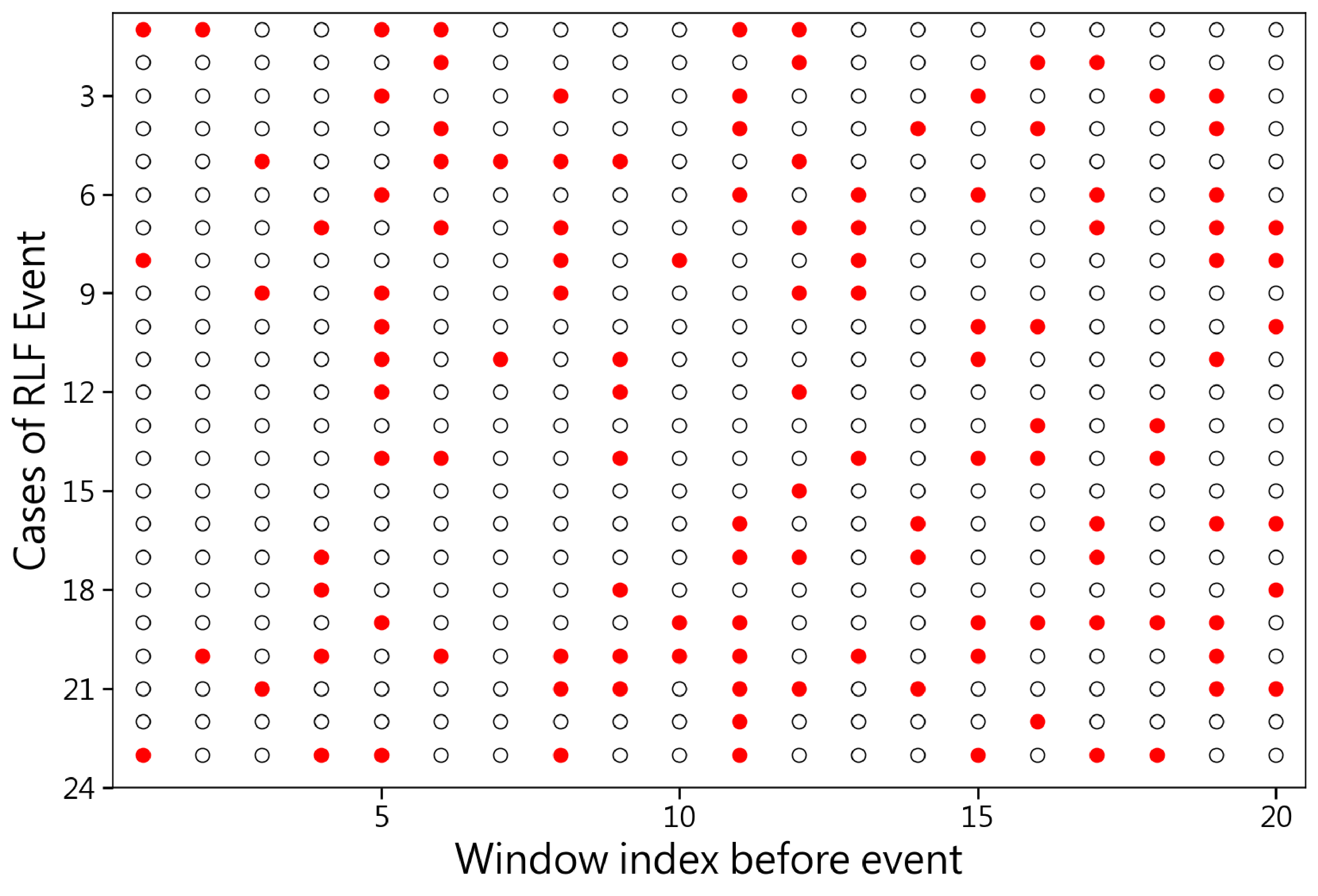}
\caption{XGBoost~\cite{Chen2016_XGBoost} with $T_p=2$~s}
\label{fig4b}
\end{subfigure}
\hfill
\begin{subfigure}{0.31\textwidth}
\centering
\includegraphics[width=\textwidth]{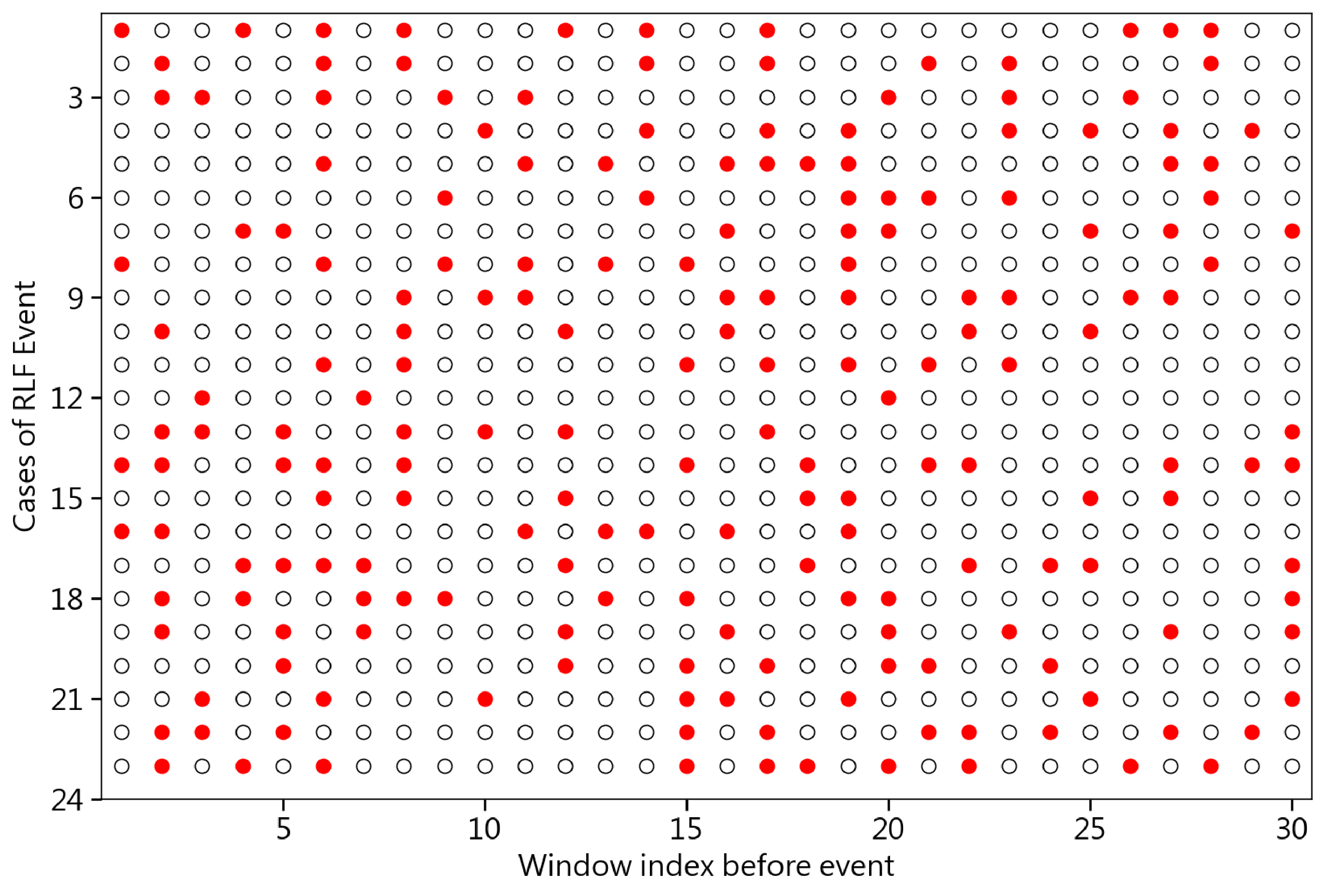}
\caption{TimesNet~\cite{Wu2023_TimesNet} with $T_p=3$~s}
\label{fig4c}
\end{subfigure}
\caption{Time-domain results of CNN, XGBoost, and TimesNet with $T_s=3$~s and $I_s=0.1$~s. Each column corresponds to a ground-truth breakdown event, and red markers indicate prediction points that correctly anticipate the corresponding event within the specified prediction horizon.}
\vspace{-0.2in}
\label{fig4}
\end{figure*}

\subsection{Implementation Environment and Model Setup}

All experiments were implemented in Python~3.10 using TensorFlow and XGBoost libraries in Visual Studio Code. Model training and evaluation were performed on a 13th~Gen Intel\textsuperscript{\textregistered} Core\texttrademark~i5-13400 CPU without GPU acceleration.

The CNN, LSTM, Anomaly Transformer, PatchTST, and TimesNet models were trained using the Adam optimizer with an initial learning rate of $1\times10^{-3}$, a batch size of 128, and a maximum of 60 epochs. Early stopping was applied according to validation area under the receiver operating characteristic curve (AUC) with a patience of 10 epochs, and ReduceLROnPlateau reduced the learning rate to a minimum of $1\times10^{-5}$. Class-weighted loss functions were used to mitigate class imbalance. XGBoost was trained using its default tree-based optimization procedure, with the class-imbalance setting described in Section~\ref{sec:measure_observation}. Detailed model structures, preprocessing procedures, and configurations are available in the project repository.\footnote{Source code and configuration files are available at: \texttt{github.com/BoneZhou/RLF-Prediction}.}

\subsection{Evaluation Setup}

The dataset was divided chronologically into training and validation subsets using a 70:30 split before sliding-window generation. Sliding windows were then constructed independently within each subset, ensuring that raw time samples were not shared across the training and validation sets. Although sliding-window sampling produces many samples, the dataset contains only 23 independent RLF-related breakdown events. A separate test split would therefore leave only a few events for validation and testing, leading to unstable event-level estimates and threshold selection. Accordingly, the results are interpreted as validation-set feasibility results rather than independent generalization estimates.

We evaluate observation windows $T_s\in\{1,2,3\}$~s and prediction horizons $T_p\in\{1,2,3\}$~s with a sampling interval of $I_s=0.1$~s. Because the positive-to-negative ratio is approximately 1:500, accuracy alone is insufficient for evaluating practical warning performance. Precision, recall, and F1 are therefore used to characterize the trade-off between missed breakdown events and false warning decisions.

\subsection{Performance Comparison}

TABLE~\ref{tab:performance3} summarizes the results for $T_s=3$~s. Increasing $T_p$ from 1~s to 2~s improves F1 for most models, whereas extending it to 3~s does not consistently provide further gains. TimesNet achieves the highest F1 of 0.8498 at $(T_s,T_p)=(3~\mathrm{s},3~\mathrm{s})$. CNN reaches a competitive F1 of 0.8208 at $T_p=2$~s, while XGBoost obtains 0.7987 at the same horizon, with precision serving as a proxy for false-alarm burden across models.

Several LSTM, Anomaly Transformer, and PatchTST configurations exhibit relatively high recall but lower precision, indicating increased sensitivity at the cost of more false warnings. Longer observation windows generally provide additional temporal context, with TimesNet benefiting most clearly from $T_s=3$~s. However, the improvement is model-dependent, suggesting that longer input sequences are useful only when the architecture can effectively exploit broader pre-event temporal patterns. 

\input{tab/performance-3s}

\subsection{Computational Cost}

Increasing $T_s$ from 1~s to 3~s raises the theoretical computational cost, but the measured per-sample inference latency increases by only approximately 0.2~ms on the evaluated CPU platform. This increase is small relative to the 1--3~s prediction horizons. End-to-end latency, including data collection, preprocessing, communication, and control action, remains for future evaluation.

\subsection{Time-Domain Early-Warning Behavior}

Fig.~\ref{fig4} illustrates the time-domain prediction behavior of CNN, XGBoost, and TimesNet using $T_s=3$~s. Correct prediction markers appear before the corresponding labeled breakdown events within the specified prediction horizon, showing that the evaluated models can generate warning decisions prior to event occurrence in the selected traces.

TABLE~\ref{tab2} evaluates several event-oriented trigger rules. The ``any one point'' rule detects all 23 considered events for the three models, whereas requiring multiple prediction points reduces event coverage. The ``any two points'' rule retains 100\% event coverage for CNN and TimesNet and 95.7\% for XGBoost. In contrast, requiring consecutive predictions is more restrictive: the two-consecutive-point rule covers 73.9\%, 65.2\%, and 69.6\% of the events for CNN, XGBoost, and TimesNet, respectively.

These results illustrate the trade-off between event coverage and trigger persistence. Since TABLE~\ref{tab2} does not quantify event-level false alarms, selection of a trigger policy also requires false-alarm evaluation.

\input{tab/eval}

\subsection{Error Analysis and Limitations}

Qualitative inspection suggests that some false negatives coincide with abrupt degradation, whereas some false positives coincide with transient changes in neighboring-cell dominance. These observations remain preliminary.

The evaluation is limited to one metro line, one operator, 23 breakdown events, and commercial smartphones rather than dedicated railway equipment with potentially different RF and antenna characteristics. Thus, the results should be interpreted as a feasibility benchmark rather than a general performance guarantee. Future work will consider larger multi-route datasets, event-level false-alarm analysis, dedicated railway hardware, and event-disjoint training, validation, and testing.

\section{Conclusion}

This paper presented a measurement-driven benchmark for seconds-ahead early warning of RLF-related reliability breakdowns in a commercial 5G NSA railway environment. Six representative learning models were evaluated using serving- and neighboring-cell measurements under different observation windows and prediction horizons. TimesNet achieved the highest F1 of 0.8498, demonstrating the predictive value of lightweight radio measurements for proactive resilience. Broader cross-network and hardware generalization remains for future investigation.

\end{document}

%% file: tab/performance-3s.tex
\begin{table}[h]
\centering
\caption{Performance comparison with observation window $T_s = 3$ s and sampling rate $I_s = 0.1$ s (10 Hz).}
\label{tab:performance3}
\setlength{\tabcolsep}{4pt}
\begin{tabular}{|c|c|c|c|c|c|}
\hline
\textbf{Model} & \textbf{$T_p$} & \textbf{Accuracy} & \textbf{Precision} & \textbf{Recall} & \textbf{F1} \\
\hline
\multirow{3}{*}{CNN~\cite{LeCun2015_NatureDL}} 
    & 1 s & 0.9806 & 0.2705 & 0.8116 & 0.4058 \\
    \cline{2-6}
    & 2 s & \textbf{0.9935} & \textbf{0.7456} & \textbf{0.9130} & \textbf{0.8208} \\
    \cline{2-6}
    & 3 s & 0.9873 & 0.6968 & 0.8551 & 0.7679 \\
\hline
\multirow{3}{*}{LSTM~\cite{Hochreiter1997_LSTM}}
    & 1 s & \textbf{0.9782} & 0.2275 & 0.6957 & 0.3429 \\
    \cline{2-6}
    & 2 s & 0.975 & 0.3870 & \textbf{0.9058} & 0.5423 \\
    \cline{2-6}
    & 3 s & 0.9781 & \textbf{0.5374} & 0.7633 & \textbf{0.6307} \\
\hline
\multirow{3}{*}{XGBoost~\cite{Chen2016_XGBoost}}
    & 1 s & \textbf{0.9962} & \textbf{0.7284} & 0.8551 & 0.7867 \\
    \cline{2-6}
    & 2 s & 0.9927 & 0.7235 & \textbf{0.8913} & \textbf{0.7987} \\
    \cline{2-6}
    & 3 s & 0.9865 & 0.6782 & 0.8551 & 0.7564 \\
\hline
\multirow{3}{*}{\shortstack{Anomaly\\Transformer~\cite{Xu2022_AnomalyTransformer}}}
    & 1 s & \textbf{0.9811} & 0.2581 & 0.6957 & 0.3765 \\
    \cline{2-6}
    & 2 s & 0.9800 & 0.4357 & \textbf{0.7609} & 0.5541 \\
    \cline{2-6}
    & 3 s & 0.9737 & \textbf{0.4762} & 0.7246 & \textbf{0.5747} \\
\hline
\multirow{3}{*}{PatchTST~\cite{Nie2023_PatchTST}}
    & 1 s & \textbf{0.9825} & 0.2690 & 0.6667 & 0.3833 \\
    \cline{2-6}
    & 2 s & 0.952 & 0.2276 & \textbf{0.8116} & 0.3556 \\
    \cline{2-6}
    & 3 s & 0.9613 & \textbf{0.3421} & 0.6280 & \textbf{0.4429} \\
\hline
\multirow{3}{*}{TimesNet~\cite{Wu2023_TimesNet}}
    & 1 s & 0.9881 & 0.3621 & 0.6087 & 0.4541 \\
    \cline{2-6}
    & 2 s & 0.9737 & 0.3735 & 0.8986 & 0.5277 \\
    \cline{2-6}
    & 3 s & \textbf{0.9924} & \textbf{0.8265} & \textbf{0.8744} & \textbf{0.8498} \\
\hline
\end{tabular}
\vspace{-0.1in}
\end{table}

%% file: tab/eval.tex
\begin{table}[h]
  \centering
  \renewcommand{\arraystretch}{1.5} 
  \caption{Performance evaluation of Fig.~\ref{fig4}.}
  \label{tab2}
  \setlength{\tabcolsep}{2pt}
  \begin{tabular}{|c|c|c|c|}
    \hline
    Index & CNN~\cite{LeCun2015_NatureDL} & XGBoost~\cite{Chen2016_XGBoost} & TimesNet~\cite{Wu2023_TimesNet} \\
    \hline
    Any one point   & $\frac{23}{23}=100\%$ & $\frac{23}{23}=100\%$ & $\frac{23}{23}=100\%$ \\
    \hline
    Any two points  & $\frac{23}{23}=100\%$ & $\frac{22}{23}=95.7\%$ & $\frac{23}{23}=100\%$ \\
    \hline
    Any three points  & $\frac{21}{23}=91.3\%$  & $\frac{20}{23}=87\%$ & $\frac{23}{23}=100\%$ \\
    \hline
    Two consecutive points & $\frac{17}{23}=73.9\%$ & $\frac{15}{23}=65.2\%$ & $\frac{16}{23}=69.6\%$ \\
    \hline
    Three consecutive points & $\frac{3}{23}=13\%$ & $\frac{3}{23}=13\%$ & $\frac{5}{23}=21.7\%$ \\
    \hline
  \end{tabular}
\vspace{-0.1in}
\end{table}